\documentclass[aps,prl,showpacs,twocolumn,superscriptaddress,
              floats,floatfix]{revtex4}
\usepackage{amssymb,amsmath}
\usepackage{graphics}
\usepackage{color}
\usepackage{subfigure}
\usepackage{epsfig}
\begin{document}
\title{Manifestations of Drag Reduction by Polymer Additives \\
	in Decaying, Homogeneous, Isotropic Turbulence}
\author{Prasad Perlekar}
\email{perlekar@physics.iisc.ernet.in}
\affiliation{Centre for Condensed Matter Theory, Department of Physics, Indian
Institute of Science, Bangalore 560012, India.} 
\author{Dhrubaditya Mitra} 
\email{Dhrubaditya.MITRA@obs-nice.fr}
\affiliation{Observatoire de la C\^ote d`Azur, BP 4229, 06304 Nice Cedex 4, 
France.}
\author{Rahul Pandit}
\email{rahul@physics.iisc.ernet.in}
\altaffiliation[\\ also at~]{Jawaharlal Nehru Centre For Advanced
Scientific Research, Jakkur, Bangalore, India.}
\affiliation{Centre for Condensed Matter Theory, Department of Physics, Indian
Institute of Science, Bangalore 560012, India.} 
\begin{abstract}
The existence of drag reduction by polymer additives, well established for 
wall-bounded turbulent flows, is controversial in homogeneous, isotropic 
turbulence. To settle this controversy we carry out a 
high-resolution direct numerical simulation (DNS) of {\it decaying}, 
homogeneous, isotropic turbulence with polymer additives. Our study 
reveals clear manifestations of drag-reduction-type phenomena: On the 
addition of polymers to the turbulent fluid we obtain a reduction in 
the energy dissipation rate, a significant modification of the fluid 
energy spectrum especially in the deep-dissipation range, a suppression 
of small-scale intermittency, and a decrease in small-scale vorticity 
filaments.
\end{abstract}
\maketitle
The dramatic reduction of drag by the addition of small 
concentrations of polymers to a turbulent fluid continues to engage 
the attention of engineers and physicists. Significant advances have 
been made in understanding drag reduction both 
experimentally \cite{vir75,dam94,too97} and 
theoretically \cite{lum73,sre00,pta03,lvo04}
in channel flows or the Kolmogorov flow \cite{bof05}. 
However, the existence of drag-reduction-type phenomena in 
turbulent flows that are homogeneous and isotropic \cite{tab86, 
bha91, ben03, ben04, kal_poly04, ang05, doo99, mcc77, fri70, 
bon93, bon05} remains controversial. Some experimental 
\cite{doo99,mcc77, fri70, bon93}, numerical 
\cite{ben03,ben04,kal_poly04,ang05}, and theoretical 
\cite{tab86,bha91} studies have suggested that drag reduction 
should occur even in homogeneous, isotropic turbulence; but other 
studies have refuted this claim \cite{bon05}.

  To settle this controversy we have initiated an extensive direct numerical 
simulation (DNS) of decaying, homogeneous, isotropic turbulence in the 
presence of polymer additives. We monitor the decay of 
turbulence from initial states in which the kinetic energy of the fluid is 
concentrated at small wave vectors; this energy then cascades down 
to large wave vectors where it is dissipated by viscous effects; the 
energy-dissipation rate $\epsilon$ attains a maximum at $t_m$, 
roughly the time at which the cascade is completed. A recent 
shell-model study \cite{kal_poly04} has suggested that this peak in 
$\epsilon$ can be used to quantify drag reduction by polymer additives. Since 
shell models are far too simple to capture the complexities of real 
flows, we have studied decaying turbulence in the 
Navier-Stokes (NS) equation coupled to the Finitely Extensible 
Nonlinear Elastic Peterlin (FENE-P) model~\cite{pet66} for polymers. 
Our study, designed specifically to uncover 
drag-reduction-type phenomena, shows that the position of the maximum in
$\epsilon$ depends only mildly on the polymer concentration $c$; however, 
the value of $\epsilon$ at this maximum falls as $c$ increases. 
We use this decrease of $\epsilon$ to define the percentage 
drag (or dissipation) reduction DR in decaying homogeneous, 
isotropic turbulence; we also explore
other accompanying physical effects and show that they are in 
qualitative accord with drag-reduction experiments~\cite{bon93,hoy77}: 
In particular, DR increases with $c$ (upto $25\%$ in one of our simulations). 
For small values of $c$ the energy spectrum of the fluid is 
modified appreciably only in the dissipation range; however, 
this suffices to yield significant drag reduction. 
We show that vorticity filaments and intermittency 
are reduced at small spatial scales and that the extension 
of the polymers decreases as $c$ increases. 
  
The NS and FENE-P (henceforth NSP) equations are 
\begin{eqnarray}
D_t{\bf u} &=& \nu \nabla^2 {\bf u}+
              \frac{\mu}{\tau_P}\nabla.[f(r_P){\cal C}] - {\nabla}p;  
                                                 \label{ns}\\
D_t{\cal C}&=& {\cal C}. (\nabla {\bf u}) + 
                {(\nabla {\bf u})^T}.{\cal C} - 
                \frac{{f(r_P){\cal C} }- {\cal I}}{\tau_P}.
                                                   \label{FENE}
\end{eqnarray}
Here ${{\bf u}({\bf x},t)}$ is the fluid velocity at point 
${\bf x}$ and time $t$, incompressibility is enforced by 
$\nabla .{\bf u}=0$, $D_t=\partial_t + {\bf u}.\nabla$, 
$\nu$ is the kinematic viscosity of the fluid, $\mu$ the viscosity 
parameter for the solute (FENE-P), $\tau_P$ the polymer relaxation time, 
$\rho$ the solvent density (set to $1$), $p$  
the pressure, $(\nabla {\bf u})^T$ the transpose of $({\nabla {\bf u}})$, 
${\cal C}_{\alpha\beta}\equiv 
             {\langle{R_\alpha}{R_\beta}\rangle}$ the 
elements of the polymer-conformation tensor ${\cal C}$ (angular 
brackets indicate an average over polymer configurations), 
${\cal I}$ the identity tensor with elements $\delta_{\alpha \beta}$, 
$f(r_P)\equiv{(L^2 -3)/(L^2 - r_P^2)}$ the FENE-P potential that ensures 
finite extensibility, $ r_P \equiv \sqrt{Tr(\cal C)}$ and $ L $ the
length and the maximum possible extension, respectively, of the polymers, 
and $c\equiv\mu/(\nu+\mu)$ a dimensionless measure of the polymer 
concentration \cite{vai03}. $c=0.1$ corresponds, roughly, 
to $100$ppm for polyethylene oxide \cite{vir75}.

We consider homogeneous, isotropic, turbulence, so we use periodic
boundary conditions and solve Eq.~(\ref{ns}) by using a 
massively parallel pseudospectral code~\cite{vin91} with $N^3$ collocation  
points in a cubic domain (side ${\mathbb L}=2\pi$). We eliminate aliasing 
errors \cite{vin91} by the 2/3 rule, to obtain reliable data at small length 
scales, and use a second-order, slaved Adams-Bashforth 
scheme for time marching. For Eq.~(\ref{FENE}) we use an 
explicit sixth-order central-finite-difference scheme in space and a 
second-order Adams-Bashforth method for temporal evolution. The numerical 
error in $r_P$ must be controlled by choosing a small time step $\delta t$, 
otherwise $r_P$ can become larger than $L$, which leads to a numerical 
instability; this time step is much smaller than what is necessary 
for a pseudospectral DNS of the NS equation alone. Table~\ref{table:para} 
lists the parameters we use. We preserve the symmetric-positive-definite 
(SPD) nature of $\cal C$ at all times by using\cite{vai03} the following 
Cholesky-decomposition scheme: If we define  
${\cal J} \equiv f(r_P) {\cal C}$, ~Eq.~(\ref{FENE}) becomes 
\begin{equation} 	
D_t{\cal J} = {\cal J}. (\nabla {\bf u}) 
+ ({\nabla \bf u})^T .{\cal J} -s({\cal J} - {\cal I})+ q {\cal J},
\label{conj} 
\end{equation} 
where $s=(L^2 -3+ j^2)/(\tau_P L^2)$,
$q=[d/(L^2 -3)-(L^2 -3+ j^2)(j^2 -3)/(\tau_P L^2(L^2 -3))]$,
$j^2\equiv Tr({\cal J})$,
and 
$d = Tr[ {\cal J}. (\nabla{\bf u}) + (\nabla{\bf u})^T .{\cal J}].$
Since ${\cal C}$ and hence ${\cal J}$ are SPD matrices, we can write
${\cal J}= {\cal LL}^T$, where ${\cal L}$ is a 
lower-triangular matrix with elements 
$\ell_{ij}$, such that  $\ell_{ij}=0$ for $j>i$.
Thus Eq.\eqref{conj} now yields $(1\le i \le3$ and $ 
\Gamma_{ij}=\partial_i u_j )$  
\begin{eqnarray}
\nonumber
{D_t \ell_{i1}}  &=& \sum_k \Gamma_{ki}\ell_{k1}
+ \frac{1}{2}\Big[(q-s)\ell_{i1}+(-1)^{(i \bmod 1)}
\frac{s\ell_{i1}}{\ell^2_{11}}\Big] \\    
\nonumber    
&&{}+ (\delta_{i3}+\delta_{i2})\frac{\ell_{i2}}{\ell_{11}}
 \sum_{m>1}\Gamma_{m1}\ell_{m2} \\
\nonumber
&&{} + \delta_{i3}\Gamma_{i1}\frac{\ell^2_{33}}{\ell_{11}},~
\mbox{for}~i\geq1; \\
\nonumber
{ D_t \ell_{i2}}  &=& \sum_{m\geqslant2}\Gamma_{mi}\ell_{m2}
 -\frac{\ell_{i1}}{\ell_{11}}
  \sum_{m\geqslant2}\Gamma_{m1}\ell_{m2}\\
\nonumber
&&{} + \frac{1}{2}\Big[(q-s)\ell_{i2}+(-1)^{(i+2)}
 s\frac{\ell_{i2}}{\ell_{22}^2}
\Big(1+\frac{\ell^2_{21}}{\ell^2_{11}}\Big)\Big]\\
\nonumber          
&&{}+ \delta_{i3}\Big[\frac{\ell^2_{33}}{\ell_{22}}
\Big(\Gamma_{32}-\Gamma_{31}\frac{\ell_{21}}{\ell_{11}}\Big)
 + s\frac{\ell_{21}\ell_{31}}{\ell^2_{11}\ell_{22}}\Big],~
\mbox{for}~i\geq2;\\
\nonumber 
{ D_t \ell_{33}} &=& \Gamma_{33} \ell_{33}
   - \ell_{33}\Big[\sum_{m<3}\frac{\Gamma_{3m}\ell_{3m}}{\ell_{mm}}\Big] 
     + \frac{\Gamma_{31}\ell_{32}\ell_{21}\ell_{33}}{\ell_{11}\ell_{22}} \\
   \nonumber
    &&{}-s\frac{\ell_{21}\ell_{31}\ell_{32}}{\ell^2_{11}\ell_{22}\ell_{33}}
       + \frac{1}{2}\Big[(q-s)\ell_{33} \\
    &&{}   + \frac{s}{\ell_{33}}  
        \Big(1+\sum_{m<3}\frac{\ell^2_{3m}}{\ell^2_{mm}}\Big)
       +\frac{s\ell^2_{21}\ell^2_{32}}{\ell^2_{11}\ell^2_{22}\ell_{33}} \Big].
\label{ellij}
\end{eqnarray}
The SPD nature of  $\cal C$ is preserved by Eq.\eqref{ellij}
if $\ell_{ii} > 0$, which we enforce explicitly~\cite{vai03} by 
considering the evolution of $\ln(\ell_{ii})$ instead of $\ell_{ii}$.

We use the following initial conditions (superscript $0$): 
${\cal C}^0_{mn}({\bf x}) =  \delta_{mn}$  for all ${\bf x}$; and
${u}^0_m({\bf k})= P_{mn}({\bf k}){v}^0_n({\bf k}) 
\exp(i{\theta_n(\bf{k})})$, 
with $m,n=x,y,z$, $P_{mn}=(\delta_{mn}-k_mk_n/k^2)$ the transverse 
projection operator, ${\bf k}$ the wave-vector with components 
$k_m= (-N/2,-N/2+1,\ldots,N/2)$,  
$k=|{\bf k}|$, $\theta_n({\bf k})$ random numbers 
distributed uniformly between $0$ and $2\pi$, and $v^0_n({\bf k})$ 
chosen such that the initial kinetic-energy spectra are 
either of type $I$, with $E^I(k) = k^2 \exp(-2{k}^4)$, or of type $II$, with 
$E^{II}(k) = k^4 \exp(-2{k}^2)$.   

In addition to ${\bf u}({\bf x},t)$, its Fourier transform 
${\bf u}_{\bf k}(t)$, and ${\cal C}({\bf x},t)$ we monitor
the vorticity $\omega \equiv \nabla \times {\bf u}$, the 
kinetic-energy spectrum 
$E(k,t)\equiv\sum_{k-1/2< k' \le k+1/2}|{\bf u}^2_{\bf k'}(t)|$,
the total kinetic energy 
~${\mathcal E}(t) \equiv\sum_kE(k,t)$, 
the energy-dissipation-rate 
$\epsilon(t) \equiv \nu \sum_k k^2 E(k,t)$, 
the cumulative probability distribution of scaled polymer extensions 
$P^C(r_P^2/L^2)$, and the hyperflatness 
${\mathcal F}_6(r) \equiv {\mathcal S}_6(r)/{\mathcal S}_2^3(r)$, 
where ${\mathcal S}_p(r) \equiv \langle\{[{\bf u}({\bf x + r})-{\bf u}({\bf x})]\cdot{\bf r}/r\}^p \rangle$ is the order-$p$ longitudinal 
velocity structure function and the angular brackets denote an 
average over our simulation domain at $t_m$.  For notational 
convenience, we do not display the dependence on $c$ explicitly.

\begin{figure*}
\begin{minipage}[t]{0.34\linewidth}
\includegraphics[width=\linewidth]{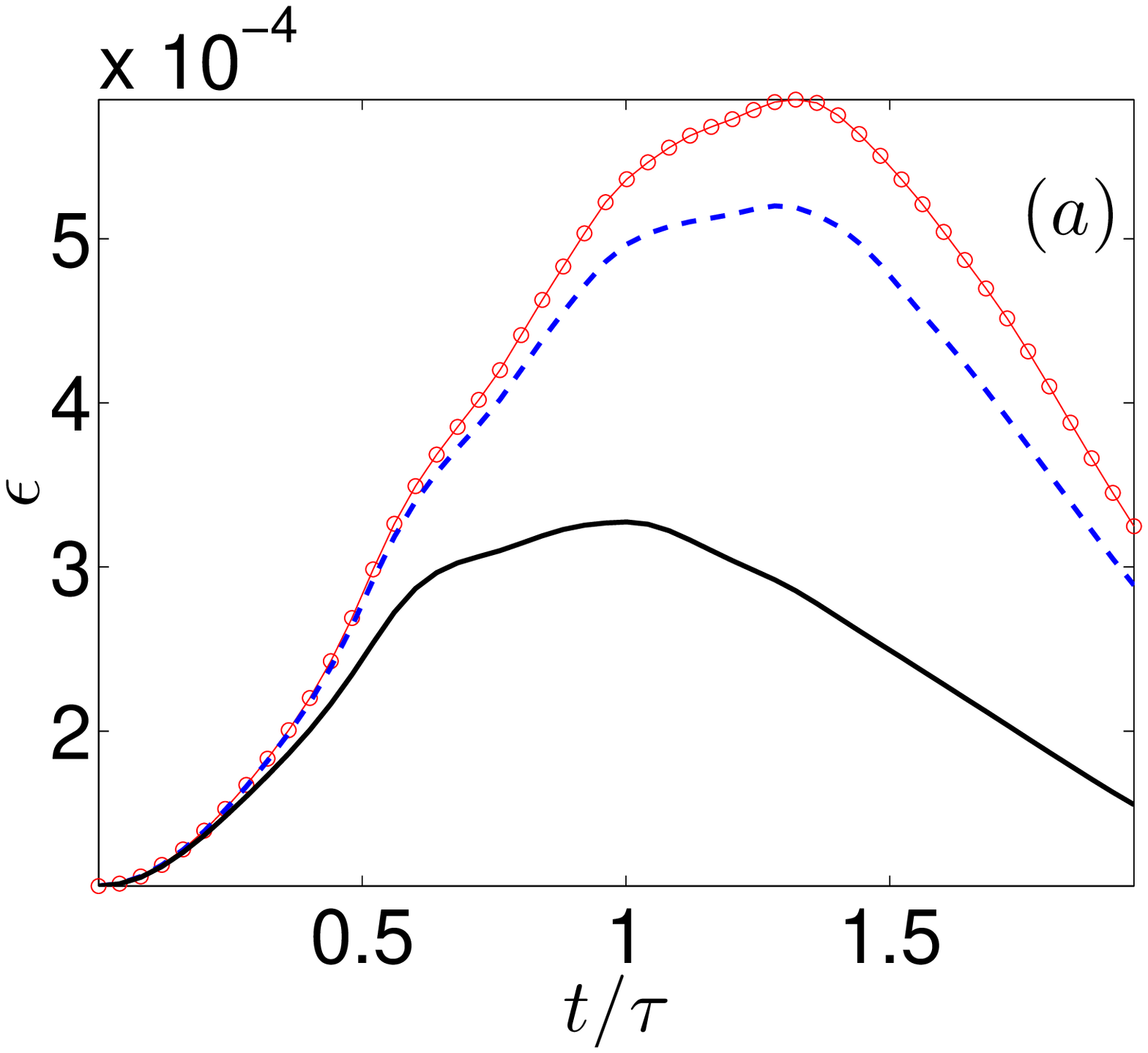}
\end{minipage} \hfill
\begin{minipage}[t]{0.31\linewidth}
\includegraphics[width=\linewidth]{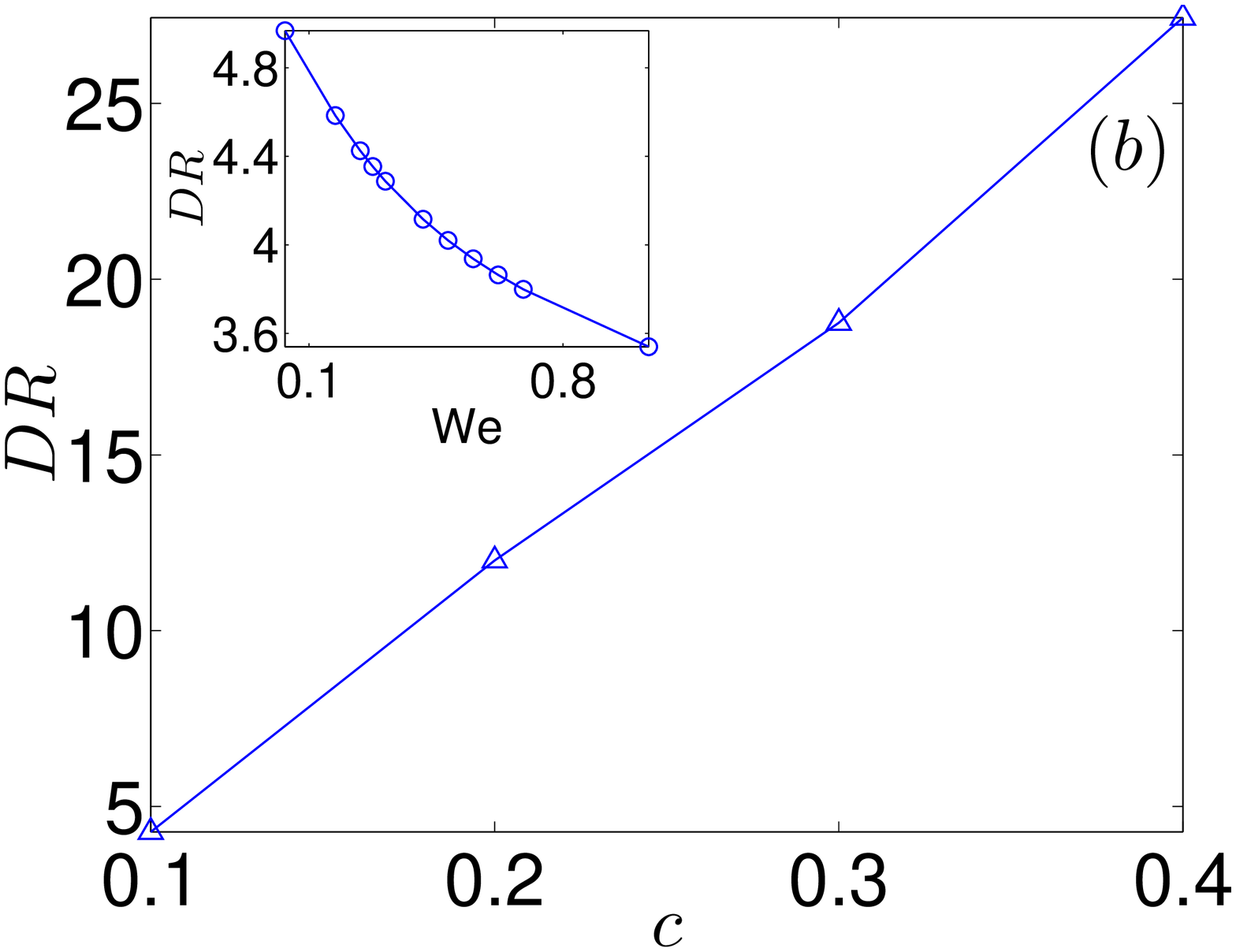}
\end{minipage} \hfill
\begin{minipage}[t]{0.32\linewidth}
\includegraphics[width=\linewidth]{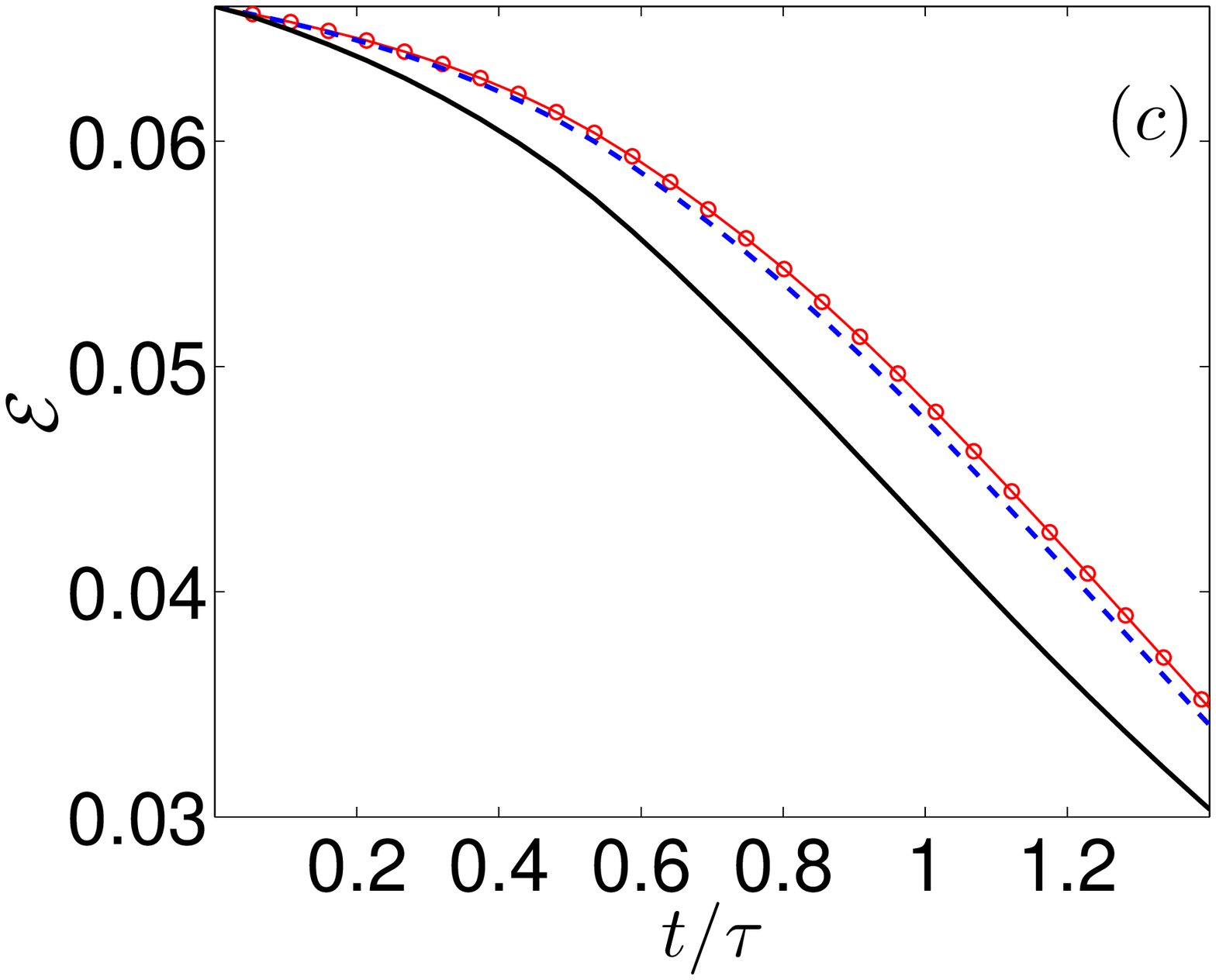}
\end{minipage}   
\caption{\small(Color online)
(a) Temporal evolution of the energy dissipation rate $\epsilon$ 
(run {\tt NSP-256}) for concentrations $c = 0.1({\textcolor{blue}{- -}})$ 
and $c = 0.4($solid line$)$, with 
$\tau\equiv\sqrt{2{\mathcal E}(t=0)/3{\mathbb L}^2}$; 
(b) percentage drag-reduction DR versus $c$ (run {\tt NSP-192}); the 
inset shows the mild variation in DR with $We$ (runs {\tt NSP-96}); 
(c) temporal evolution of the total fluid energy ${\mathcal E}$ 
for concentrations $c = 0.1({\textcolor{blue}{- -}})$ and 
$c=0.4($solid line$)$ (runs {\tt NSP-256}). In (a) and (c) 
the plots for $c=0$ $(\textcolor{red}{o -})$ are shown for comparison.}
\label{eddr}
\end{figure*}

Figure~(\ref{eddr}a) shows that $\epsilon$ first increases
with time, reaches a peak, and then decreases; for
$c=0$ this peak occurs at $t = t_m$. The position of this peak 
changes  mildly with $c$ but its height goes down 
significantly as $c$ increases. This suggests the following 
natural definition~\cite{kal_poly04} of the 
percentage drag or dissipation reduction for decaying homogeneous, 
isotropic turbulence:    
\begin{eqnarray}
\begin{aligned}
{\rm DR}\equiv\left(\frac{\epsilon^{f,m}-\epsilon^{p,m}}
{\epsilon^{f,m}}\right)\times 100;
\end{aligned}
\label{dragreduction}
\end{eqnarray}
here (and henceforth) the superscripts $f$ and $p$ stand, respectively, 
for the fluid without and with polymers and the superscript $m$ 
indicates the time $t_m$.  Figure~(\ref{eddr}b) shows plots of 
${\rm DR}$ versus $c$, for the Weissenberg number 
$We \equiv \tau_P\sqrt{\epsilon^{f,m}/\nu} \simeq 0.35$, 
and versus $We$, for $c = 1/11 \simeq 0.1$.  
${\rm DR}$ increases with $c$ in qualitative accord with experiments 
on channel flows (where ${\rm DR}$ is defined via a normalized pressure 
difference); but it drops  gently as $We$ increases, in contrast to the 
behavior seen in channel flows (in which $\tau_P$ is varied by changing 
the polymer).

\begin{figure*}
\begin{minipage}[t]{0.32\linewidth}
\includegraphics[width=\linewidth]{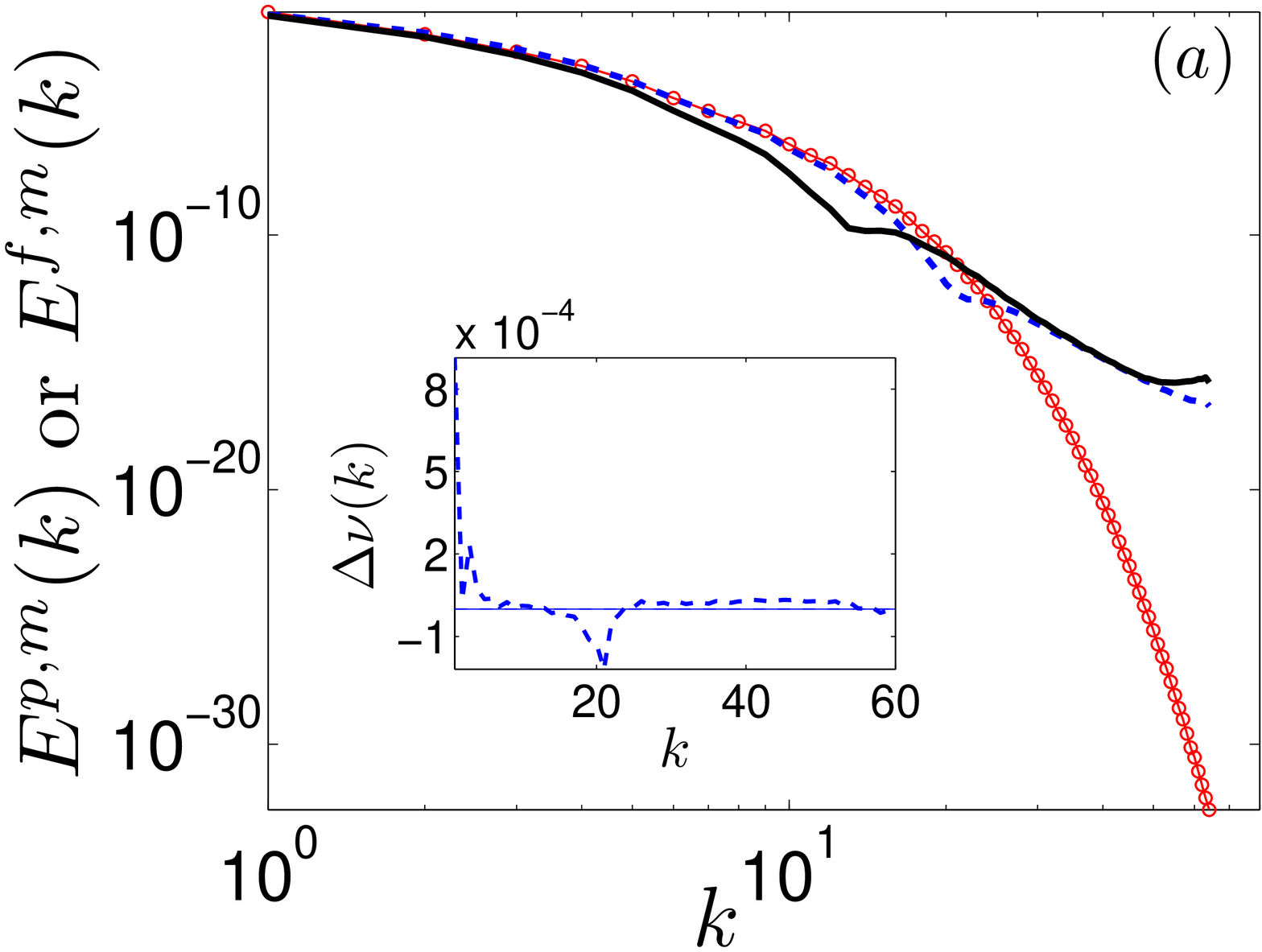}
\end{minipage}\hfill
\begin{minipage}[t]{0.32\linewidth}
\includegraphics[width=\linewidth]{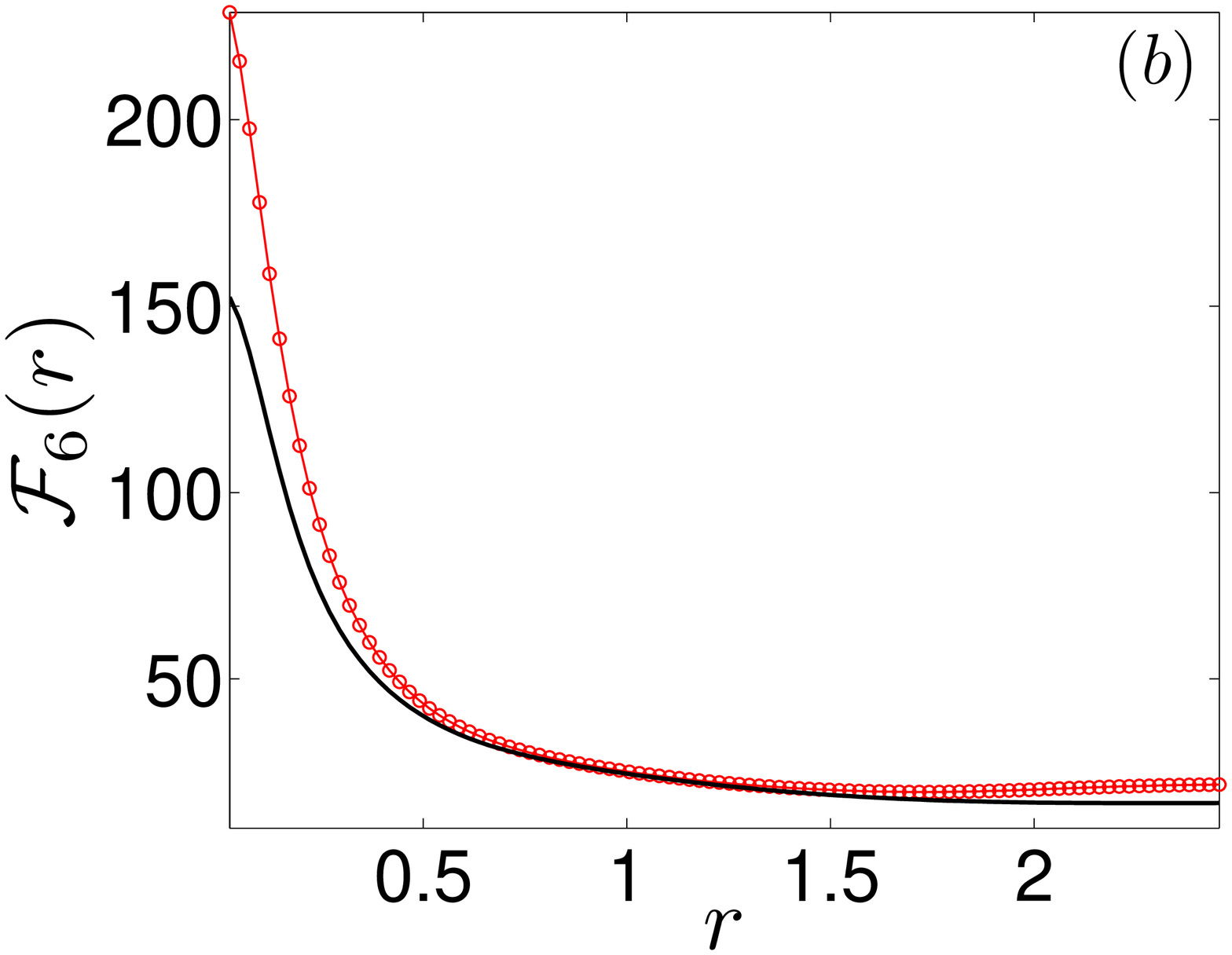}
\end{minipage} \hfill
\begin{minipage}[t]{0.32\linewidth}
\includegraphics[width=\linewidth]{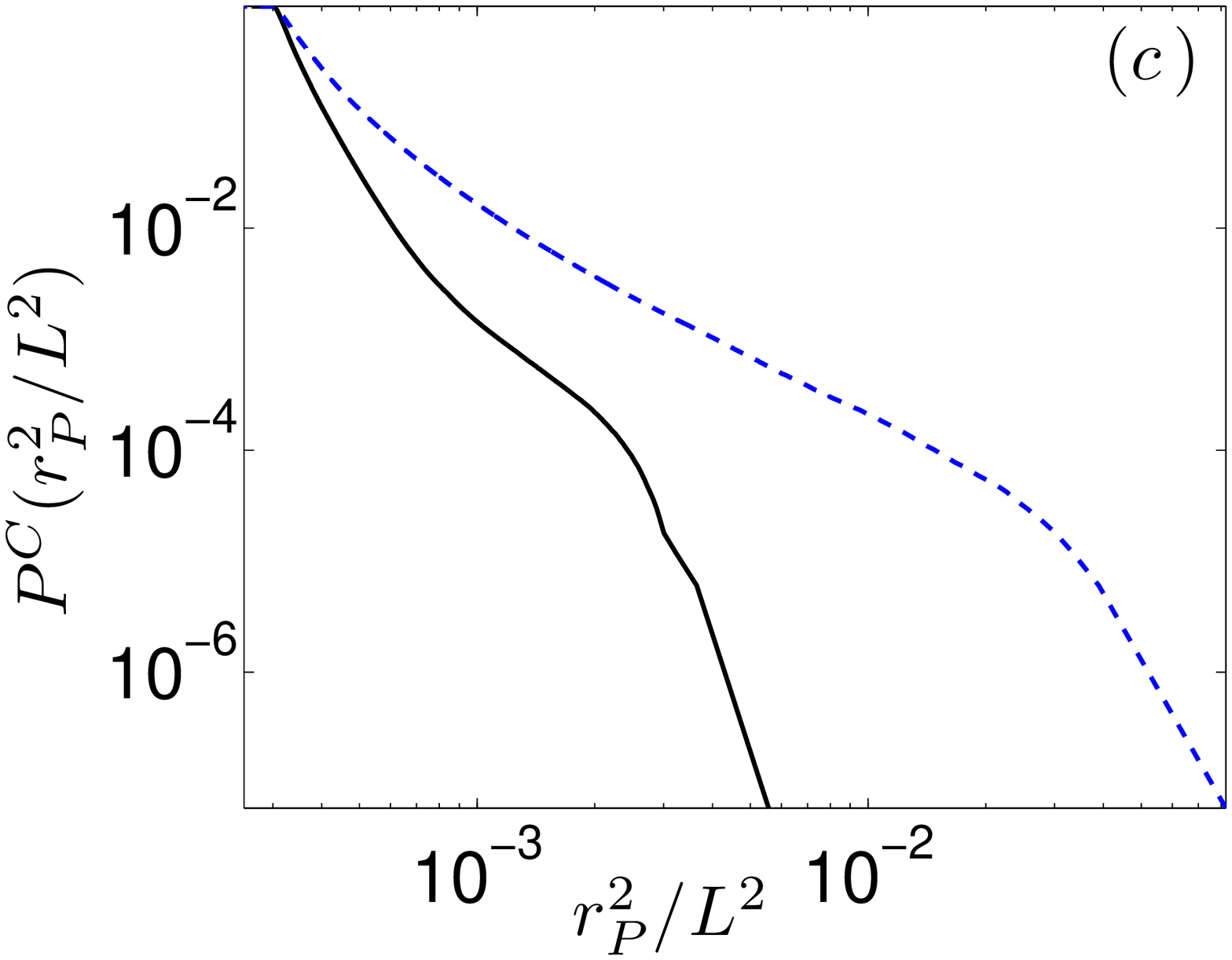}
\end{minipage}
\caption{\small(Color online) 
(a) Plots of the energy spectra $E^{p,m}(k)$ or $E^{f,m}(k)$ versus $k$ 
(run {\tt NSP-192}) for $c = 0.1({\textcolor{blue}{- -}})$ and 
$c=0.4($solid line$)$ [$E^{p,m}(k)$ is unchanged if we use 
$N=256$, with all other parameters the same (run {\tt NSP-256A})]; 
inset: polymer contribution to the scale-dependent 
viscosity $\Delta \nu(k)$ versus $k$ for $c=0.1(\textcolor{blue}{--})$; 
$\Delta \nu(k)=0$ $($solid line$)$ is also shown for reference; 
(b) the hyper-flatness ${\mathcal F}_6(r)$ as a function of $r$ 
(run {\tt NSP-256}) and concentration $c = 0.4($solid line$)$. In (a) and (b) 
the corresponding plots with $c=0$ $(\textcolor{red}{o-})$ are shown 
for comparison.  (c) The cumulative PDF $P^C(r_P^2/L^2)$ versus $r_P^2/L^2$ 
for $c=0.1(\textcolor{blue}{- -})$ and $c=0.4($solid line$)$ 
(run {\tt NSP-256}). 
} 
\label{spec}
\end{figure*}

In decaying turbulence, the total kinetic energy 
${\mathcal E}(t)$ of the fluid falls as $t$ increases; the rate at 
which it falls increases with $c$ [Fig.~(\ref{eddr}c)], 
which suggests that the addition of polymers 
increases the effective viscosity of the solution. This is not  
at odds with the decrease of $\epsilon$ with increasing $c$  
since the effective viscosity because of polymers turns out to 
be {\it scale-dependent}.  We confirm this by obtaining 
the kinetic-energy spectrum $E^{p,m}(k)$ for the fluid in the 
presence of polymers  at $t=t_m$. For small concentrations 
($c \simeq 0.1$) the spectra with and without 
polymers differ substantially only in the deep dissipation range, 
where $E^{f,m}(k)\ll E^{p,m}(k)$. As $c$ increases, to say $c\simeq0.4$, 
$E^{p,m}(k)$ is reduced relative to $E^{f,m}(k)$ at intermediate values 
of $k$ [Fig.~(\ref{spec}a)]; however, deep in the  dissipation range 
$E^{f,m}(k)\ll E^{p,m}(k)$. We now define \cite{ben04} 
the effective scale-dependent viscosity $\nu_e(k)\equiv \nu+\Delta \nu(k)$, 
with $\Delta \nu(k)\equiv -\mu \sum_{k-1/2 < k' \le k+1/2} {\bf u}_{\bf k'} 
\cdot (\nabla \cdot {\cal J})_{\bf -k'}/[\tau_P{k'}^2E^{p,m}(k')]$, where 
$(\nabla \cdot {\cal J})_{\bf k}$ is the Fourier transform of 
$\nabla \cdot {\cal J}$. The inset of 
Fig.~(\ref{spec}a) shows that $\Delta \nu(k)>0$ for $k<15$, but 
$\Delta\nu(k)<0$ around $k=20$. This explains why $E^{p,m}(k)$ is 
suppressed relative to $E^{f,m}(k)$ at small $k$, rises above it 
in the deep-dissipation range, and crosses over from its small-$k$ 
to large-$k$ behaviors around the value of $k$ where $\Delta\nu(k)$ 
goes through zero.

\begin{figure}
\begin{minipage}[t]{0.45\linewidth}
\includegraphics[width=\linewidth]{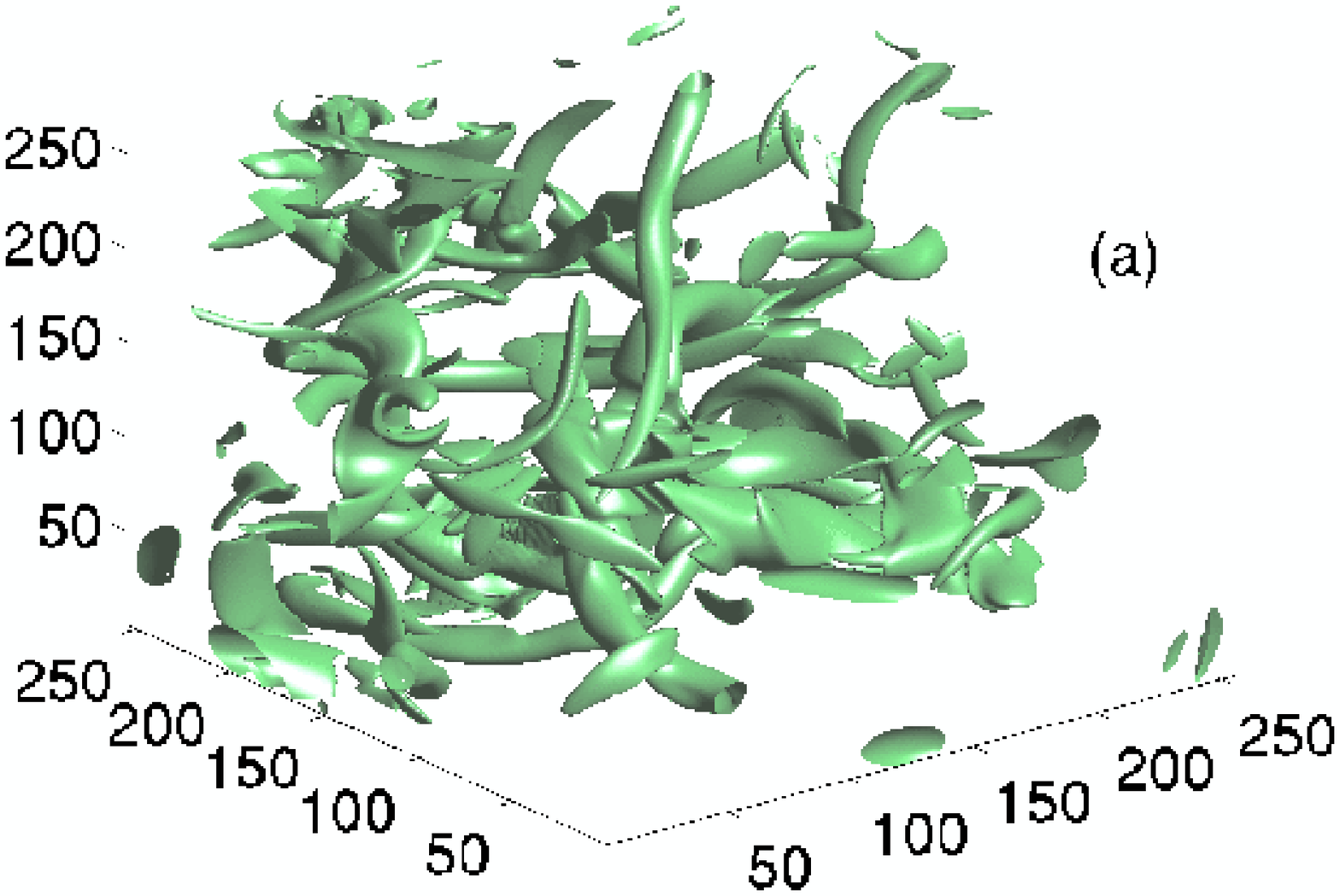}
\end{minipage} \hfill
\begin{minipage}[t]{0.45\linewidth}
\includegraphics[width=\linewidth]{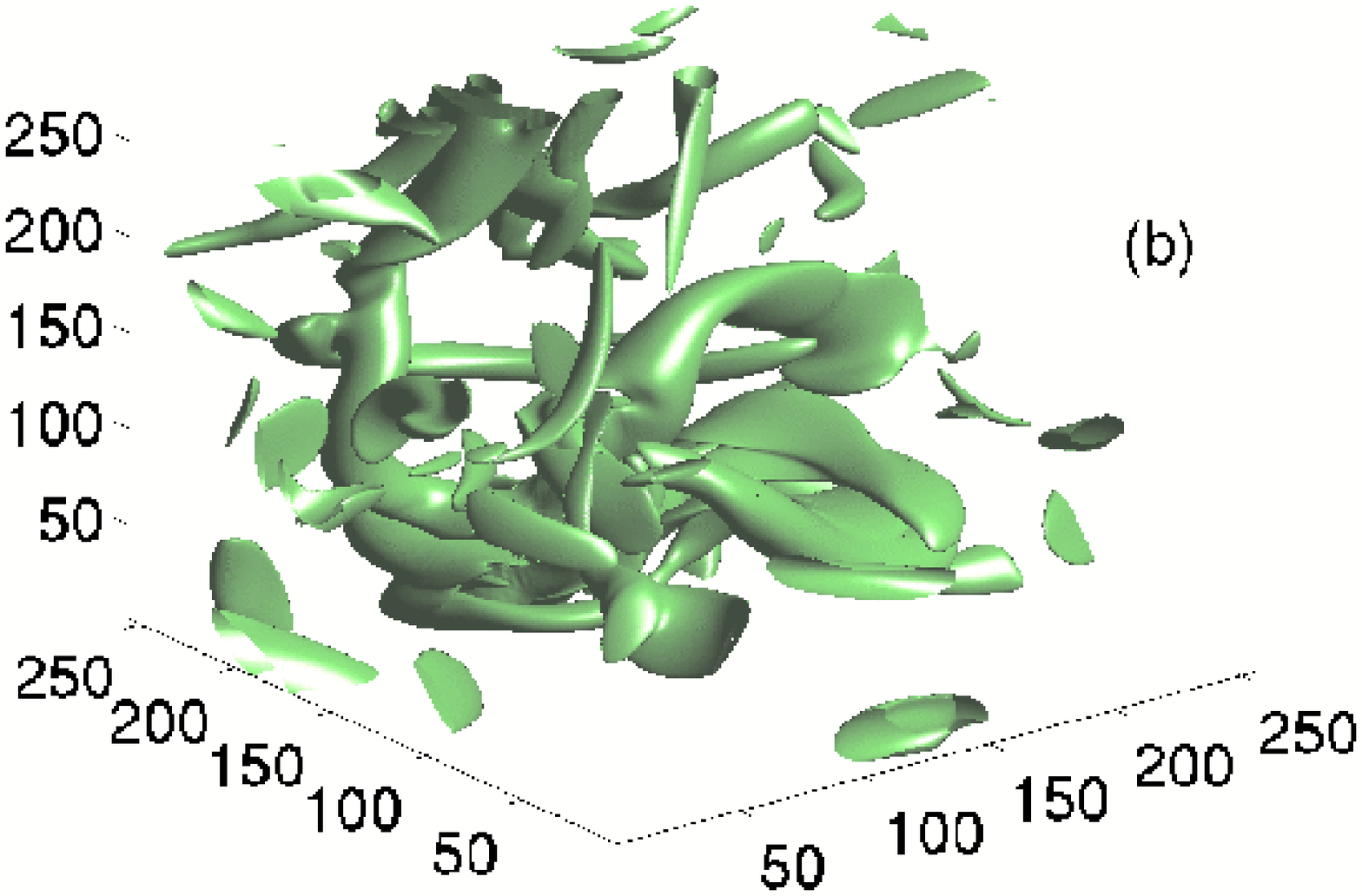}
\end{minipage} \hfill
\caption{\small(Color online)
Constant-$|\omega|$ isosurfaces for 
$|\omega|=\langle{|\omega|}\rangle+2\sigma$ at $t_m$ 
without (a) and with (b) polymers, (run {\tt NSP-256}) 
and  $c = 0.4$;  $\langle|\omega|\rangle$ is the mean  
and $\sigma$ the standard deviation of $|\omega|$.
} 
\label{vor}
\end{figure}

Given the resolution of our DNS, inertial-range intermittency can be 
studied only by using extended self similarity \cite{ben93} as we will 
report elsewhere. However, we explore dissipation-range statistics 
further by calculating the hyperflatness ${\mathcal F}_6(r)$ 
[Fig.~(\ref{spec}b)].  The addition of polymers slows 
down the  growth of ${\mathcal F}_6(r)$, as $r\rightarrow0$, which signals 
the reduction of small-scale intermittency. This  
is further supported by the iso-$|\omega|$ surfaces shown in 
Fig.~(\ref{vor}).  If no polymers are present, these iso-$|\omega|$ 
surfaces are filamentary \cite{kan03} for large $|\omega|$; 
polymers suppress a significant fraction of these filaments.
 
We use a rank-order method \cite{mit05a} to obtain $P^C(r_P^2/L^2)$ 
and find that, as $c$ increases [Fig.~(\ref{spec}c)], 
the extension of the polymers decreases. We have checked that, in the 
passive-polymer version of Eqs.\eqref{ns} and \eqref{FENE}, the 
extension of polymers is much more than in Fig.~(\ref{spec}c). 

Our study contrasts clearly drag reduction in homogeneous, 
isotropic, turbulence and in wall-bounded flows. In both 
these cases the polymers increase the overall viscosity 
of the solution (see, e.g., Fig.~(\ref{eddr}c) and Ref.\cite{ben04}). 
In wall-bounded flows the presence of  polymers inhibits the flow of 
the stream-wise component of the momentum into the wall, which, in turn, 
increases the net throughput of the fluid and thus results in drag reduction,
a mechanism that can have no analog in homogeneous, isotropic turbulence. 
However, the decrease of $\epsilon(t)$ with increasing $c$ [Fig.~(\ref{eddr}b)] 
yields a natural definition of DR [Eq.\eqref{dragreduction}] for this case 
\footnote{In some steady-state simulations \cite{vai03,ben03} DR is 
associated with $E^p(k)>E^f(k)$, for small $k$. We obtain this for  
type $II$, but not type $I$, initial conditions; but 
Eq.\eqref{dragreduction} yields drag reduction for both of these 
initial conditions.}. 
Thus, if the term {\it drag reduction} must be reserved for wall-bounded 
flows, then we suggest the expression {\it dissipation reduction} for 
homogeneous, isotropic, turbulence. We have shown that $\nu_e$ must be 
scale-dependent; its counterpart in wall-bounded flows is 
the position-dependent viscosity of Refs.~\cite{lum73,lvo04}. 
Furthermore, as in wall-bounded flows, an increase in $c$ leads to an 
increase in DR  [Fig.~(\ref{eddr}b)]. In channel flows an increase 
in $We$ leads to an increase in DR, but we find that DR falls marginally 
as $We$ increases [Fig.~(\ref{eddr}b)].

Our DNS of the Navier-Stokes equation with polymer 
additives [Eqs. \eqref{ns} and \eqref{FENE}] resolves the controversy 
about drag reduction in decaying homogeneous, isotropic turbulence and shows 
clearly that Eq.~\eqref{dragreduction} offers a natural definition of DR for 
this case in a far more realistic model than those of 
Refs.~\cite{kal_poly04,ben03}. We also find a nontrivial modification 
of the fluid kinetic-energy spectrum especially in the deep-dissipation 
range [Fig.~(\ref{spec}b)] that can be explained in terms of a 
polymer-induced, scale-dependent viscosity. Experiments 
\cite{fri70,mcc77} do not resolve the dissipation range as clearly as 
we do, so the experimental verification of the deep-dissipation-range 
behavior of Fig.~(\ref{spec}a) remains a challenge. Earlier theoretical 
studies \cite{bha91,ben03} have also not concentrated on this 
dissipation range. The reduction in the small-scale intermittency 
[Fig.~(\ref{spec}b)] and in the constant-$|\omega|$ isosurfaces 
[Fig.~(\ref{vor})] is in qualitative agreement with channel-flow  
studies \cite{dam94}, where a decrease in the turbulent volume fraction 
is seen on the addition of the polymers, and water-jet studies \cite{hoy77}, 
where the addition of the polymers leads to a decrease in small-scale 
structures. We hope our work will stimulate more experimental 
studies of drag or dissipation reduction in homogeneous, isotropic turbulence.  
\begin{table}
   \begin{tabular}{@{\extracolsep{\fill}} c c c c c c c}
    \hline
    $ $ &$N$ & $\delta{t}$ & $L$ & $\nu$ & $\tau_P$ & $c$ \\
   \hline \hline
    {\tt NSP-96}  & $96$  &  $1.0\times10^{-2}$ & $100$ &  $10^{-2}$  & $0.1-3$ & $~0.1,~0.2,~0.3,~0.4$ \\
    {\tt NSP-192} & $192$ &  $1.0\times10^{-2}$ & $100$ &  $10^{-2}$  &    $1$  & $~0.1,~0.4$ \\
    {\tt NSP-256A} & $256$ &  $1.0\times10^{-2}$ & $100$ &  $10^{-2}$ &    $1$  & $~0.1,~0.4$ \\
    {\tt NSP-256} & $256$ &  $4.0\times10^{-3}$ & $100$ &  $10^{-3}$ &    $1$  & $~0.1,~0.4$ \\
\hline
\end{tabular}
\caption{\small
The parameters $N$, $\delta t$, $L$, $\nu$, $\tau_P$ and $c$ 
for our four runs ${\tt NSP-96}$, ${\tt NSP-192}$, ${\tt NSP-256A}$, and ${\tt NSP-256}$. 
${\tt NSP-96}$, ${\tt NSP-192}$, ${\tt NSP-256A}$ use type $I$ initial conditions;   
${\tt NSP-256}$ uses an initial condition of type $II$. We also 
carry out DNS studies of the NS equation with the same numerical resolutions 
as our NSP runs. $Re\equiv\sqrt{20}{\mathcal E^{f,m}}/\sqrt{3\nu\epsilon^{f,m}}$ and 
$We \equiv \tau_P\sqrt{\epsilon^{f,m}/\nu}$;  
{\tt NSP-96}: $Re=47.1$ and $We=0.03, 0.17, 0.24, 
0.28, 0.31, 0.41, 0.48, 0.55, 0.62, 0.68, 1.03$; {\tt NSP-192} and 
{\tt NSP-256A}: $Re=47.1$ and $We=0.35$; {\tt NSP-256}: $Re=126.6$ and $We=0.76$.
}
\label{table:para}
\end{table} 

We thank C.~Kalelkar, R.~Govindarajan, V.~Kumar, S.~Ramaswamy, L.~Collins,  
and A.~Celani for discussions, CSIR, DST, and UGC(India) for financial 
support, and SERC(IISc) for computational facilities. DM is supported 
by the Henri Poincar\'e Postdoctoral Fellowship.


\begin{thebibliography}{10}

\bibitem{vir75}
P. Virk, AIChE {\bf 21},  625  (1975).

\bibitem{dam94}
P. van Dam, G. Wegdam, and J. van~der Elsken, J. Non-Newtonian Fluid Mech. {\bf
  53},  215  (1994).

\bibitem{too97}
J.~D. Toonder, M. Hulsen, G. Kuiken, and F. Nieuwstadt, J. Fluid Mech. {\bf
  337},  193  (1997).

\bibitem{lum73}
J. Lumley, J. Polym. Sci {\bf 7},  263  (1973).

\bibitem{sre00}
K. Sreenivasan and C. White, J. Fluid Mech. {\bf 409},  149  (2000).

\bibitem{pta03}
P. Ptasinski {\it et~al.}, J.~Fluid Mech {\bf 490},  251  (2003).

\bibitem{lvo04}
V. L'vov, A. Pomyalov, I. Procaccia, and V. Tiberkevich, Phys. Rev. Lett. {\bf
  92},  244503  (2004).

\bibitem{bof05}
G. Boffetta, A. Celani, and A. Mazzino, Phys.~Rev.~E {\bf 71},  036307  (2005).

\bibitem{tab86}
M. Tabor and P.~D. Gennes, Europhys. Lett. {\bf 2},  519  (1986).

\bibitem{bha91}
J.~K. Bhattacharjee and D. Thirumalai, Phys. Rev. Lett. {\bf 67},  196  (1991).

\bibitem{ben03}
R. Benzi, E. de~Angelis, R. Govindarajan, and I. Procaccia, Phys.~Rev.~E {\bf
  68},  016308  (2003).

\bibitem{ben04}
R. Benzi, E. Ching, and I. Procaccia, Phys.~Rev.~E {\bf 70},  026304  (2004) 
consider a scale-dependent viscosity for a shell model (but use an artificial 
diffusivity for polymers for numerical stability).

\bibitem{kal_poly04}
C. Kalelkar, R. Govindarajan, and R. Pandit, Phys.~Rev. E {\bf 72},  017301
  (2004).

\bibitem{ang05}
E. de~Angelis, C. Casicola, R. Benzi, and R. Piva, J.~Fluid~Mech {\bf 531},  1
  (2005).

\bibitem{doo99}
E. van Doorn, C. White, and K. Sreenivasan, Phys. Fluids {\bf 11},  2387
  (1999).

\bibitem{mcc77}
W. McComb, J. Allan, and C. Greated, Phys.~Fluids {\bf 20},  873  (1977).

\bibitem{fri70}
C. Friehe and W. Schwarz, J.~Fluid~Mech {\bf 44},  173  (1970).

\bibitem{bon93}
D. Bonn, Y. Couder, P. van Dam, and S. Douady, Phys.~Rev.~E {\bf 47},  R28
  (1993).

\bibitem{bon05}
D. Bonn {\it et~al.}, J.~Phys.~CM {\bf 17},  S1219  (2005).

\bibitem{pet66}
A. Peterlin, J.~Polym.~Sci., Polym.~Lett. {\bf 4},  287  (1966); H. Warner,
  Ind. Eng. Chem. Fundamentals {\bf 11}, 379 (1972); R. Armstrong, J. Chem.
  Phys. {\bf 60} 724 (1974); E. Hinch, Phys. Fluids {\bf 20}, S22 (1977).

\bibitem{hoy77}
J. Hoyt and J. Taylor, Phys. Fluids {\bf 20},  S253  (1977).

\bibitem{vai03}
T. Vaithianathan and L. Collins, J. Comput. Phys. {\bf 187},  1  (2003). We
  correct their Eq.(40) and definition of $q$.

\bibitem{vin91}
A. Vincent and M. Meneguzzi, J.~Fluid~Mech. {\bf 225},  1  (1991); C. Canuto,
  M. Hussaini, A. Quarteroni, and T. Zang, {\em Spectral Methods in 
Fluid Dynamics} (Spinger-Verlag, Berlin, 1988).

\bibitem{ben93}
R. Benzi {\it et~al.}, Phys. Rev. E {\bf 48},  R29  (1993); S. Dhar, A. Sain,
  and R. Pandit, Phys. Rev. Lett. {\bf 78}, 2964 (1997).

\bibitem{kan03}
Y. Kaneda {\it et~al.}, Phys. Fluids {\bf 15},  L21  (2003).

\bibitem{mit05a}
D. Mitra, J. Bec, R. Pandit, and U. Frisch, Phys. Rev. Lett. {\bf 94},  194501
  (2005).

\end{thebibliography}
\end{document}